\documentclass[twoside,english,aps,manuscript,superscriptaddress,preprint,notitlepage,11pt]{revtex4-1}
\usepackage[T1]{fontenc}
\usepackage{mathrsfs}
\usepackage{amsmath}
\usepackage{amssymb}
\usepackage{graphicx}
\usepackage{esint}

\makeatletter
\@ifundefined{textcolor}{}
{%
 \definecolor{BLACK}{gray}{0}
 \definecolor{WHITE}{gray}{1}
 \definecolor{RED}{rgb}{1,0,0}
 \definecolor{GREEN}{rgb}{0,1,0}
 \definecolor{BLUE}{rgb}{0,0,1}
 \definecolor{CYAN}{cmyk}{1,0,0,0}
 \definecolor{MAGENTA}{cmyk}{0,1,0,0}
 \definecolor{YELLOW}{cmyk}{0,0,1,0}
}

\usepackage[hypertexnames=false]{hyperref}
\hypersetup{colorlinks=true,citecolor=blue,linkcolor=blue,urlcolor=blue} 

\AtBeginDocument{
\renewcommand{\ref}[1]{\autoref{#1}}

}
\makeatother

\usepackage{babel}
\begin{document}

\title{Anisotropic Gilbert damping in perovskite La$_{0.7}$Sr$_{0.3}$MnO$_{3}$
thin film\smallskip{}
}

\author{Qing Qin}

\affiliation{Department of Materials Science and Engineering, National University
of Singapore, Singapore 117575}

\author{Shikun He{*}}

\affiliation{Data Storage Institute, Agency for Science, Technology and Research
(A{*}STAR), 2 Fusionopolis Way 08-01 Innovis, Singapore 138634}

\email{heshikun@gmail.com msecj@nus.edu.sg}

\author{Haijun Wu}

\affiliation{Department of Materials Science and Engineering, National University
of Singapore, Singapore 117575}

\author{Ping Yang}

\affiliation{Department of Materials Science and Engineering, National University
of Singapore, Singapore 117575}

\affiliation{Singapore Synchrotron Light Source (SSLS), National University of
Singapore, 5 Research Link, Singapore 117603}

\author{Liang Liu}

\affiliation{Department of Materials Science and Engineering, National University
of Singapore, Singapore 117575}

\author{Wendong Song}

\affiliation{Data Storage Institute, Agency for Science, Technology and Research
(A{*}STAR), 2 Fusionopolis Way 08-01 Innovis, Singapore 138634}

\author{Stephen John Pennycook}

\affiliation{Department of Materials Science and Engineering, National University
of Singapore, Singapore 117575}

\author{Jingsheng Chen{*}}

\affiliation{Department of Materials Science and Engineering, National University
of Singapore, Singapore 117575}

\email{msecj@nus.edu.sg}

\begin{abstract}
The viscous Gilbert damping parameter governing magnetization dynamics
is of primary importance for various spintronics applications. Although,
the damping constant is believed to be anisotropic by theories. It
is commonly treated as a scalar due to lack of experimental evidence.
Here, we present an elaborate angle dependent broadband ferromagnetic
resonance study of high quality epitaxial La$_{0.7}$Sr$_{0.3}$MnO$_{3}$
films. Extrinsic effects are suppressed and we show convincing evidence
of anisotropic damping with twofold symmetry at room temperature.
The observed anisotropic relaxation is attributed to the magnetization
orientation dependence of the band structure. In addition, we demonstrated
that such anisotropy can be tailored by manipulating the stain. This
work provides new insights to understand the mechanism of magnetization
relaxation.
\end{abstract}
\maketitle

\section{introduction}

The magnetization relaxation process determines the speed of magnetization
relaxation and the energy required for current-induced magnetization
reversal \citep{Slonczewski_STT_1996,Katine_STT_GMR,Ikeda_2010_Namat,Ralph2008,Brataas_ST_2012,FukamiShapeMTJ}.
Understanding the mechanism and controlling of magnetization relaxation
\citep{Brataas2008a,spinPumping_PRL,Schoen2016,heinrich_2011_YIG_4_cavity,Okada2017,damping_breathMode_PRL},
including intrinsic Gilbert damping and extrinsic effects, pave the
way for ultra-low power and high performance spintronic devices based
on spin transfer and spin orbit torques \citep{STT_block_2014,Urazhdin2014,SOT_YIG2017}.
It has been demonstrated that Gilbert damping constant ($\alpha$)
can be tuned effectively by engineering the density of states and
spin orbit coupling (SOC) \citep{Schoen2016,Kubota2009,PhysRevLett.110.077203,2017LowDamping}.
In addition, magnetization relaxations subjected to finite size and
interfacial effects have also been extensively investigated \citep{spinPumping_PRL,Moriyama2008,ModeSizeDependent}.
However, it is still an open question that if magnetic damping is
anisotropic. In principle, $\alpha$ is magnetization orientation
dependent and should be a 3$\times$3 tensor in the phenomenological
Gilbert equation \citep{Gilbert1955,Steiauf2005}, yet it is often
treated as a scalar (isotropic). In the case of polycrystalline thin
films prepared by sputtering, such treatment is reasonable due to
the smearing of long range structural order. Whereas for single crystal
thin films, it is still difficult to draw a conclusion due to the
lack of convincing experimental evidence. From the view of theories,
the Gilbert damping is determined by two scattering processes, the
interband resistivity-like scattering and the intraband conductivity-like
scattering \citep{damping_breathMode_PRL}. Both terms vary with temperature
through their dependence on electron relaxation time. The interband
scattering which dominates damping in most ferromagnets becomes isotropic
at room temperature \citep{Gilmore2010}. Therefore, anisotropic linewidth
in 3d magnetic metals was only observed at low temperature\citep{Rudd1985}.
From the aspect of experimental technique, Seib et al. have predicted
that the precession trajectory of magnetization in a ferromagnetic
resonance (FMR) measurement (standard technique for measuring damping)
may partially average out the anisotropy \citep{Seib2009}. Hence,
detecting the anisotropy in Gilbert damping is extremely difficult.
Furthermore, the existence of several angle dependent extrinsic contributions
to damping in most materials further hinders the determination of
a possible weak anisotropic damping \citep{PhysRevB.69.184417,Arias-PRB-1999,rippleTMS,Okada2017}.
We note that in a ferromagnet with nearly half-metallic band structure,
the isotropic interband term is suppressed \citep{Butler_lowDamping}
and the damping can be dominated by the anisotropic intraband contribution\citep{Gilmore2010}.
Recent reports have claimed the observation of anisotropic damping
in half-metallic Heusler alloy\citep{Kasatani2014,Yilgin2007}. However,
unavoidable chemical disorder \citep{Wen2014,PhysRevB.96.224425}of
Heusler alloy introduces extrinsic effects such as spin wave scattering
hence complicates the verification procedure of such anisotropy. 

La$_{0.7}$Sr$_{0.3}$MnO$_{3}$ (LSMO) is an oxide perovskite material
exhibited half-metallic band structure and ultra-low damping at room
temperature \citep{LSMO_halfMetal_nature,Qin2017}. In this work,
we studied the magnetization relaxation of LSMO films deposited on
$\textrm{NdGaO}{}_{3}$ (NGO) (110) substrates using angle-resolved
broadband ferromagnetic resonance. The purpose of choosing NGO (110)
substrates is to utilize its non-equal $a$ and $b$ axis value. Such
asymmetry will potentially lead to non-spherical Fermi surface. Two
types of high quality samples with different static magnetic anisotropies
were investigated. The normal LSMO film (hereafter denoted as S-LSMO)
exhibited weak uniaxial magnetic anisotropy whereas the other with
modulated strain relaxation mode (hereafter denoted as W-LSMO) have
both uniaxial and cubic anisotropy fields. The angle dependence of
the in-plane intrinsic Gilbert damping showed two-fold symmetry in
both type of samples. Strikingly, the orientation of minimum damping
differs 90 degree. This work provided strong evidence of anisotropic
nature of magnetization relaxation and demonstrated the tuning of
anisotropy in damping through stress relaxation engineering.

\section{Results}

\subsection*{Epitaxial growth of LSMO}

Pulsed laser deposition (PLD) was used to deposit LSMO thin films
with a thickness of 25$\,$nm on (110) NGO substrates. The energy
and repetition frequency of KrF laser (248$\,$nm) were 225$\,$mJ
and 2$\,$Hz, respectively. During deposition, the substrate temperature
was fixed at 950$^{\circ}C$. The oxygen pressure was 225$\,$mTorr for S-LSMO and 200$\,$mTorr for W-LSMOAfter deposition,
S-LSMO was cooled down to room temperature at 10K/min under the oxygen
pressure of 1 Torr, whereas W-LSMO at 5$\,$K/min under the oxygen
pressure of 100 Torr in order to promote the modification of strain
hence micro-structurestructure.

\subsection*{Crystalline quality analysis}

\begin{figure}
\centering{}\includegraphics[width=14cm]{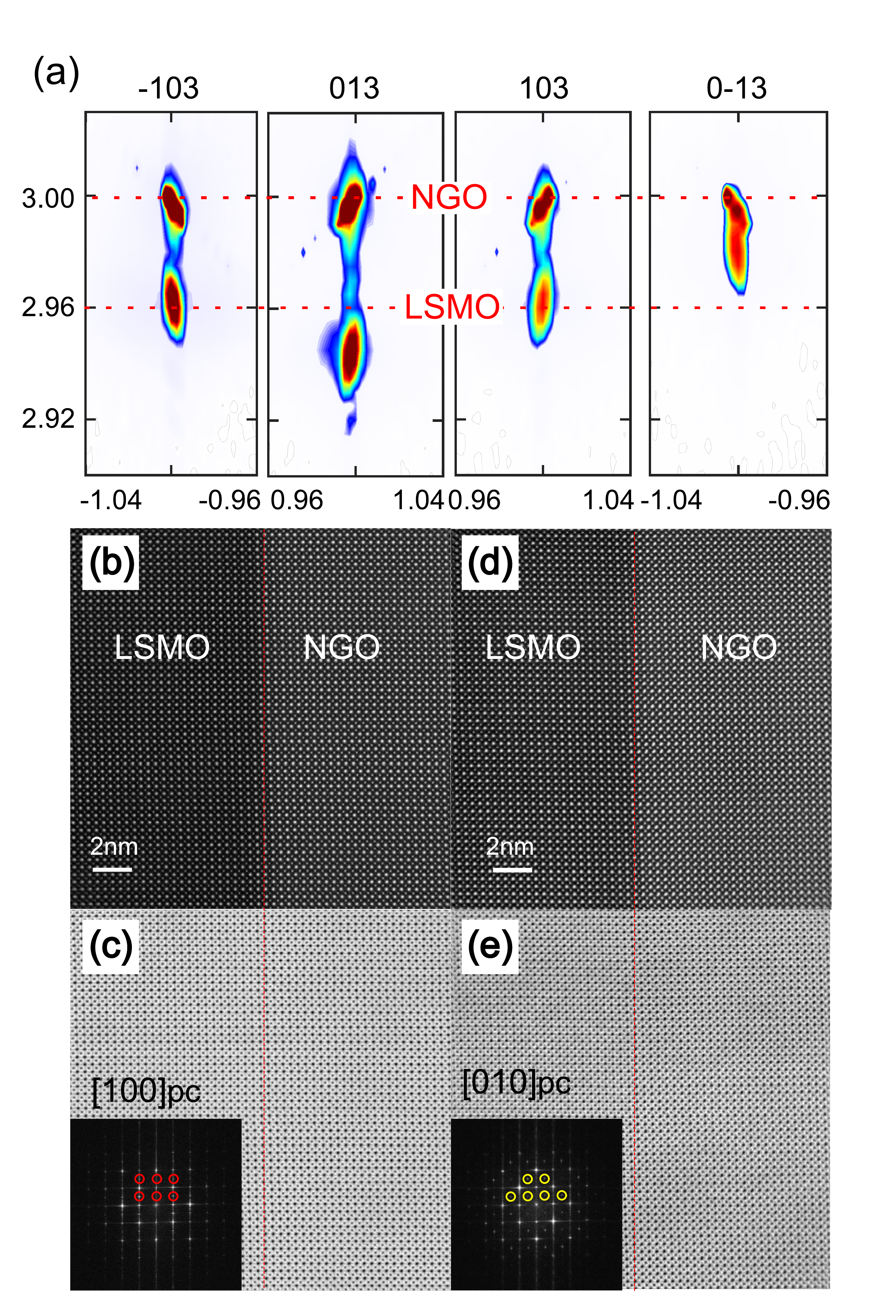}\protect\caption{\textbf{Structure characterization of S-LSMO sample.} (a) and (b)
XRD profiles around S-LSMO (00L) reflections (L=1,2,3,4) with the
incident beam aligned along the {[}100{]}$_{\textrm{pc}}$ and {[}010{]}$_{\textrm{pc}}$
, respectively. (b) and (c) STEM-HAADF/ABF lattice images of S-LSMO
along {[}100{]}$_{\textrm{pc}}$ direction. (d) and (e) STEM-HAADF/ABF
images of S-LSMO along {[}010{]}$_{\textrm{pc}}$ direction. the insets
are the intensity profile and FFT image; The red dashed line indicates
the interface.\label{fig:XRD_Main}}
\end{figure}

The crystallographic structures of the films were characterized by
synchrotron high resolution X-ray diffraction. Reciprocal space maps
(RSMs) taken at room temperature around \{013\}$_{\textrm{pc}}$ (here
the subscript pc stands for pseudocubic) reflections confirm the epitaxial
growth of LSMO layers on the NGO substrate as shown in \ref{fig:XRD_Main}
(a). The vertical alignment of LSMO and NGO reciprocal lattice point
clearly shows that the LSMO film is completely strained on the NGO
substrate. Lattice mismatch along {[}100{]}$_{\textrm{pc}}$ and {[}010{]}$_{\textrm{pc}}$
are 1.03\% and 0.8\%, respectively. Considering the position of the
LSMO reciprocal lattice point in the \{013\}$_{\textrm{pc}}$ mappings,
equal $L$ values of (103)$_{\textrm{pc}}$  and (-103)$_{\textrm{pc}}$
indicates the perpendicular relation between vector $a$ and $c$
in the lattice, whereas different L values for (013)$_{\textrm{pc}}$
and (0-13)$_{\textrm{pc}}$ shows that the angle between $b$ and
$c$ is not equal to 90°. Thus, the LSMO is monoclinic phase which
is consistent with previous reports \citep{Vailionis2011}. The good
crystalline quality was further verified by aberration-corrected scanning
transmission electron microscopy (AC-STEM). \ref{fig:XRD_Main} (b,
c) are the simultaneously acquired high angle annular dark field (HAADF)
and annular bright field (ABF) images of S-LSMO along {[}100{]}$\textrm{pc}$
direction, while \ref{fig:XRD_Main} (d, e) are for {[}010{]}$_{\textrm{pc}}$
direction. The measurement directions can be differentiated from the
diffraction of NGO substrate: 1/2{[}010{]} superlattices for {[}100{]}$_{\textrm{pc}}$
direction (inset of \ref{fig:XRD_Main}(c)) and 1/2{[}101{]} superlattices
for {[}100{]}$_{\textrm{pc}}$ direction (inset of \ref{fig:XRD_Main}(e)).
High quality single crystalline films are essential for the present
purposes because high density of defects will result in spin wave
scattering \citep{PhysRevB.69.184417}.

\subsection*{Magnetic anisotropy fields}

\begin{figure}
\begin{centering}
\includegraphics[width=14cm]{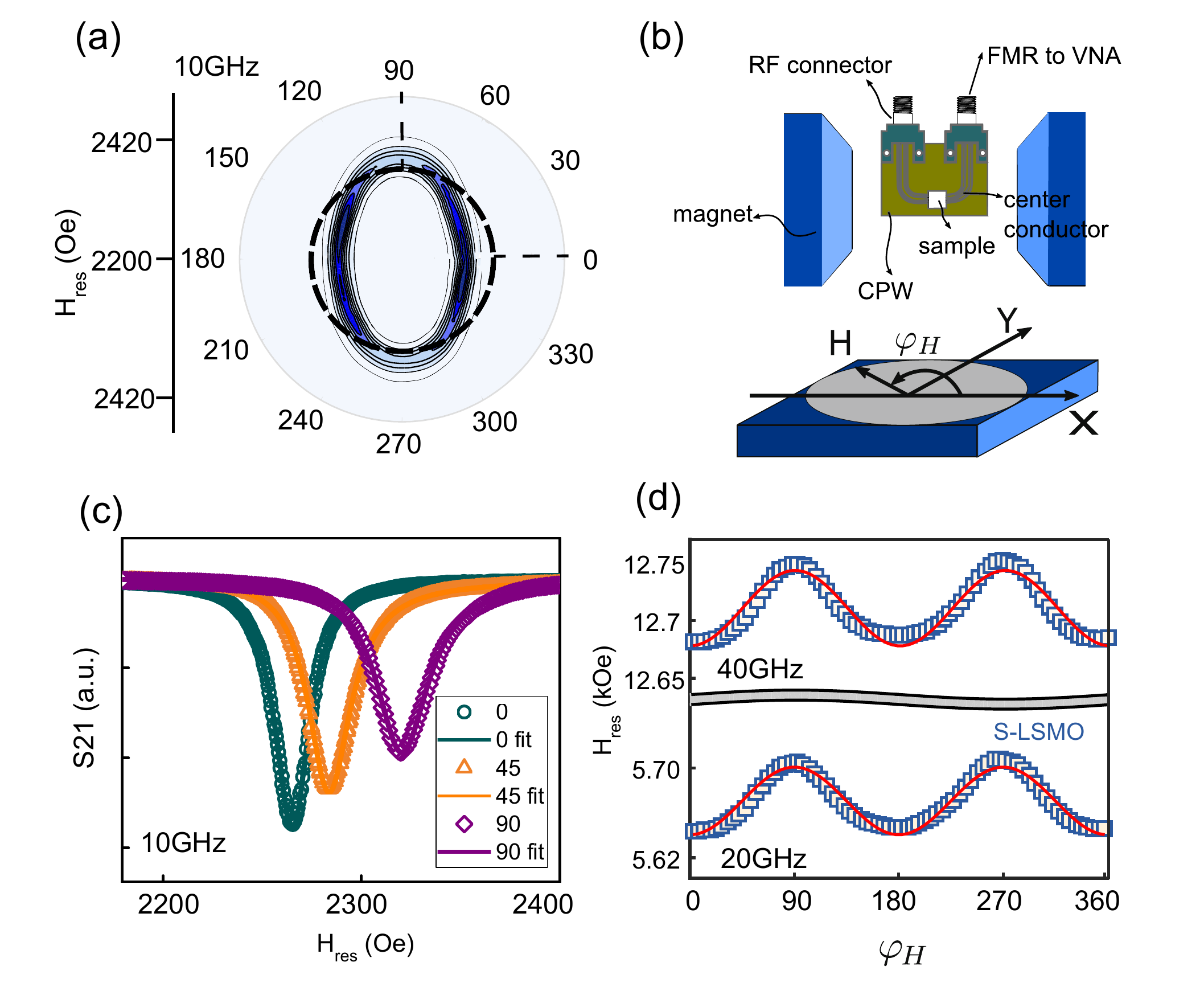}
\par\end{centering}

\protect\caption{\textbf{Magnetic anisotropy characterization.} (a) The 2D polar color
plot of the FMR spectra of S-LSMO. The frequency is 10GHz. (b) Schematics
of the FMR setup and the definition field orientation. (c) FMR spectra
for $\varphi_{H}$=0, 45 and 90 degrees for S-LSMO. (d) Field orientation
($\varphi_{H}$) dependence of the resonance fields ($H_{\textrm{res}}$)
of the S-LSMO sample at f=20 and 40GHz. The solid lines in (c) and
(d) are calculated values. \label{fig:Main_anisotropy}}
\end{figure}

The magnetic dynamic properties were investigated by a home-built
angle-resolved broadband FMR with magnetic field up to 1.5$\,$T.
All measurements were performed at room temperature. Shown in \ref{fig:Main_anisotropy}(a)
is the color-coded plot of the transmission coefficient S21 of the
S-LSMO sample measured at 10$\,$GHz. $\varphi_{H}$ is the in-plane
azimuth angle of the external magnetic field counted from {[}010{]}$_{\textrm{pc}}$
direction (\ref{fig:Main_anisotropy}(b)). This relative orientation
was controlled by a sample mounting manipulator with a precision of
less than 0.1$^{\circ}$. The olive shape of the color region indicates
the existence of anisotropy field, whereas the very narrow field region
of resonances is an evidence of low damping. Three line cuts at $\varphi_{H}$=0,
45 and 90 degrees are plotted in \ref{fig:Main_anisotropy}(c), showing
the variation of both FMR resonance field ($\mathit{H}_{\textrm{res}}$)
and line shape with $\varphi_{H}$. All curves are well fitted hence
both the $H{}_{\textrm{res}}$ and resonance linewidth $\Delta$H
are determined. The $\varphi_{H}$ dependence of H$_{\textrm{res}}$
at two selected frequencies (20 and 40 GHz) are shown in \ref{fig:Main_anisotropy}(d)
for S-LSMO. The angle dependencies of the resonance field $H{}_{\textrm{res}}(\varphi_{H})$
is calculated starting from the total energy \citep{Farle_review_1998}:
\begin{equation}
\begin{array}{l}
E=-MH\left[\cos\theta_{H}\cos\theta_{M}{\rm +}\sin\theta_{H}\sin\theta_{M}\cos(\varphi_{M}-\varphi_{H})\right]+2\pi M^{2}\cos^{2}\theta_{M}-\frac{1}{2}MH_{2\bot}\cos^{2}\theta_{M}\\
-\frac{1}{4}MH_{4\bot}\cos^{4}\theta_{M}-\frac{1}{2}MH_{2\parallel}\sin^{2}\theta_{M}\cos^{2}(\varphi_{M}-\phi_{2IP})-\frac{1}{4}MH_{4\parallel}\frac{3+\cos4(\varphi_{M}-\phi_{4IP})}{4}\sin^{4}\theta_{M}
\end{array}\label{eq:Total_Energy}
\end{equation}

where $\theta_{M}$ and $\varphi_{M}$ are the polar angle and the
azimuth angle of the magnetization ($M$), $H_{2\bot}$, $H_{4\bot}$,
$H_{2\parallel}$, $H_{4\parallel}$ are the uniaxial and cubic out-fo-plane
and in-plane anisotropy fields. The easy axes of in-plane anisotropies
are along $\phi_{2\textrm{IP}}$ and $\phi_{4\textrm{IP}}$, respectively.
According to Smit-Beljers equation the resonance condition for $\theta_{M}=\mbox{\ensuremath{\pi}/2}$
is \citep{Smit1955}: 
\begin{equation}
2\pi f{\rm =}\frac{\gamma}{M\sin\theta}\sqrt{E_{\theta\theta}E_{\varphi\varphi}}\label{eq:FMR_peak}
\end{equation}
Here, $E_{\theta\theta}=H_{{\rm res}}\cos(\varphi_{M}-\varphi_{H})+4\pi M_{{\rm eff}}-H_{2\parallel}\cos^{2}(\varphi_{M}-\phi_{2\textrm{IP}})+H_{4\parallel}(3+\cos4(\varphi_{M}-\phi_{4\textrm{IP}})/4)$
and $E_{\varphi\varphi}=H_{{\rm res}}\cos(\varphi_{M}-\varphi_{H})+H_{2\parallel}\cos2(\varphi_{M}-\phi_{2\textrm{IP}})+H_{4\parallel}\cos4(\varphi_{M}-\phi_{4\textrm{IP}})$
are second partial derivatives of the total energy with respect to
the polar and azimuth angles. $\gamma$=1.76$\times$10$^{7}$s$^{-1}$G$^{-1}$
denotes the gyromagnetic ratio, $4\pi M_{{\rm eff}}=4\pi M-H_{2\bot}$
is the effective magnetization. The resonance field of S-LSMO shows
pronounced minimum at $\varphi_{H}=n\cdot\pi$, indicating the existence
of uniaxial magnetic anisotropy with easy axis along $\phi_{2\textrm{IP}}=0$
or {[}010{]}$_{\textrm{pc}}$ direction. Cubic anisotropy is negligible
hence $H_{4\parallel}$=0. Such uniaxial anisotropy observed in S-LSMO
is consistent with previous reports \citep{Boschker2009a}, which
is attributed to anisotropic strain produced by the NGO(110) substrate
\citep{Suzuki1998,BOSCHKER20112632,Tsui2000}. Compared to the resonance
fields in our measurement, the magnetic anisotropy fields are orders
of magnitude smaller. Therefore, the calculated difference between
$\varphi_{H}$ and $\varphi_{M}$ are always smaller than $1^{\circ}$
and $\varphi=\varphi_{H}=\varphi_{M}$ is assumed in the following
discussion.

\subsection*{Magnetization orientation dependence of Gilbert damping}

\begin{figure}
\begin{centering}
\includegraphics[width=14cm]{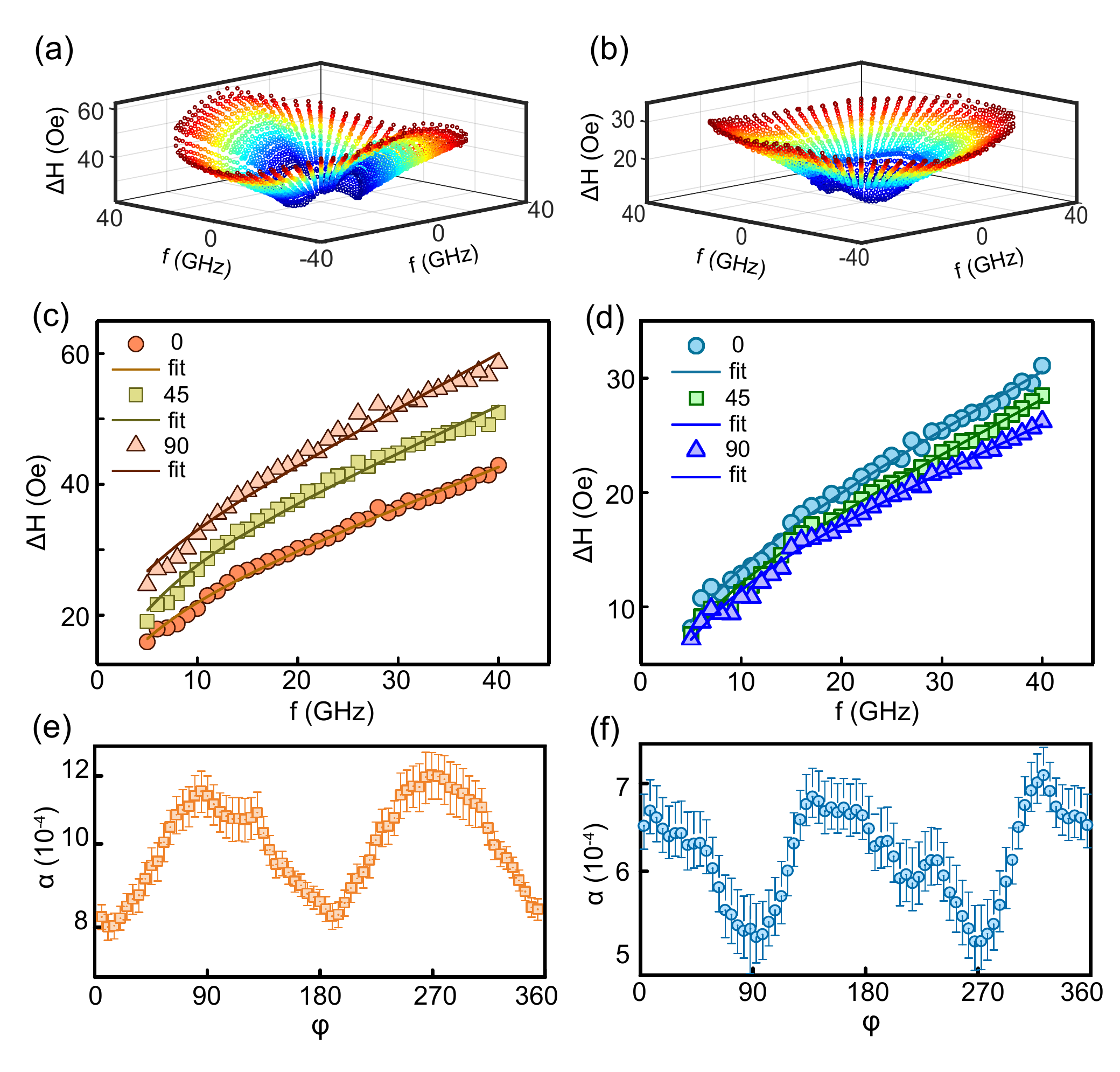}
\par\end{centering}

\protect\caption{\textbf{Anisotropic linewidth and damping:} (a)-(b) 3-D plot of frequency
and in-plane field orientation dependence of FMR linewidth. (c)-(d)
frequency dependence of FMR linewidth for seleted field orientations.
Solid symbols are experimental data and the lines are calculated value.
(e)-(f) Damping constant as a function of $\varphi$. (a),(c), (e)
are for S-LSMO and (b),(d), (f) are for W-LSMO.\label{fig:damping}}
\end{figure}

In order to study the symmetry of magnetization relaxation of the
sample. The FMR linewidth $\Delta H$ for a matrix of parameter list
(72 field orientations and 36 frequency values) are extracted. The
results are shown by 3-D plots in \ref{fig:damping}(a) . Here, $z$
axis is $\Delta H$ and $x$, $y$ axes are $f\cdot\cos\varphi$ and
$f\cdot\sin\varphi$, respectively. The figure clearly shows that
the linewidth depends on magnetization orientation. At a given frequency,
the linewidth is maximum (minimum) at $\varphi=0$ ($\varphi=\pi/2$)
for S-LSMO. \ref{fig:damping}(c) shows the $\Delta H$ versus frequency
for three field orientations. The FMR linewidth due to intrinsic magnetic
damping scales linearly with frequency ${\rm \Delta}H_{GL}=4\pi\alpha f/\gamma{\rm cos}\left(\varphi_{M}-\varphi_{H}\right)$
according to Laudau-Lifshitz-Gilbert phenomenological theory \citep{NTU_FMR,heinrich_inhomo_1991}.
However, a weak non-linearity in the low frequency range can be identified.
In general, extrinsic linewidth contributions such as inhomogeneity
and magnon scattering will broaden the FMR spectrum hence result in
additional linewidth contributions scales non-linearly with frequency
\citep{Schoen2016,Okada2017}. The interfacial magnon scattering is
suppressed due to relative large film thickness (25$\ $nm) and the
bulk magnon scattering contribution to the linewidth is negligible
in our samples with very good atomic order. However, the static magnetic
properties of the thin film may vary slightly in the millimeter scale.
Since the FMR signal is an averaged response detected by the coplanar
waveguide (5$\,$mm long), a superposition of location resonance modes
broadens the FMR spectrum. Such well-known contribution to linewidth,
defined as ${\rm \Delta}H_{{\rm inhom}}$, are generally treated as
a constant \citep{Schoen2016,heinrich_inhomo_1991,Shaw2011}. However,
it is frequency dependent for in-plane configuration and need to be
treated carefully for samples with ultra-low damping. Here, we fit
the data with ${\rm \Delta}H={\rm \Delta}H_{{\rm GL}}+{\rm \Delta}H_{{\rm inhom}}$,
taking into account the frequency and orientation dependence of ${\rm \Delta}H_{{\rm inhom}}$.
As can be seen from \ref{fig:damping}(c), the data are well reproduced
for every field orientations. Hence, the magnetization orientation
dependence of intrinsic damping constant is determined and plotted
in \ref{fig:damping}(e). Remarkably, the damping constant shows two-fold
symmetry. The lowest damping of S-LSMO with in-plane magnetization,
observed at $\varphi=0$ and $\varphi=\pi$, is $\left(8.4\pm0.3\right)\times10^{-4}$
and comparable to the value measured under a perpendicular field (\textbf{\ref{tab:SummaryValues}}).
The maximum damping at $\varphi=\pi/2$ and $\varphi=3\pi/2$ is about
25\% higher.

Since the magnetization damping and resonance field of the S-LSMO
sample exhibited identical symmetry (\ref{fig:Main_anisotropy} (d)
and \ref{fig:Main_anisotropy}(e)), it seems that the observed anisotropic
damping is directly related to crystalline anisotropy. Therefore,
we prepared the W-LSMO sample with slightly different structure and
hence modified static magnetic anisotropy properties. The W-LSMO sample
exhibited 1D long range atomic wave-like modulation \citep{Vailionis2011}
(twining domain motif) along {[}100{]}$_{\textrm{pc}}$ axis near
the interface between substrate and film. Due to different strain
relaxation mechanism as compared to S-LSMO, the $\varphi_{H}$ dependence
of $H_{\textrm{res}}$ for the W-LSMO have additional features and
can only be reproduced by including both $H_{2\parallel}(13.9\pm0.9$
Oe) and $H_{4\parallel}(11.8\pm1.2$ Oe) terms. The easy axis of the
uniaxial anisotropy ($\phi{}_{2\textrm{IP}}$=0 ) is the same as S-LSMO
whereas the additional cubic anisotropy is minimum at $\phi{}_{4\textrm{IP}}$=45°.
The magnetization orientation dependence of the FMR linewidth for
W-LSMO is significantly different (\ref{fig:damping}(b)) as compared
to S-LSMO. Such change in trend can be clearly identified from the
frequency dependence of linewidth for selected magnetization orientations
shown in \ref{fig:damping}(d). Magnetization damping values are extracted
using the same procedure as S-LSMO because the spin wave contribution
is excluded. The damping constant again showed two-fold in-plane symmetry.
However, in contrast to S-LSMO, the maximum damping value of W-LSMO
is observed at $\varphi=0$ and $\varphi=\pi$.

\section{Discussion}

Anisotropy in linewidth at low temperatures have been reported decades
ago, however, data in most early publications were taken at a fixed
frequency in a cavity-based FMR \citep{Rudd1985,Vittoria1967}. Due
to lack of frequency dependence information, it is not clear if the
anisotropy in linewidth is due to intrinsic damping or extrinsic effects
\citep{Dubowik2011,PhysRevB.84.054461,PhysRevB.58.5611}. In this
study, besides wide range of frequencies, we also adopted samples
with effective anisotropy orders of magnitude smaller than the external
field. Therefore, the field dragging effect and mosaicity broadening,
both of which are anisotropic in natur e\citep{Mosacity2007}, are
negligibly small and the Gilbert damping constant is determined reliably.
Furthermore, the mechanism in this simple system is different from
previous reports related to interfacial exchange coupling and spin
pumping\citep{LeGraet2010,BakerAnisotropy2016}. Since both S-LSMO
and W-LSMO exhibited in-plane uniaxial magnetic anisotropy, the opposite
trends observed in these two samples exclude the existence of a direct
link between anisotropic damping and effective field. Both magnetic
anisotropy and damping are related to the band structure but in quite
different ways. According to perturbation theory, the magnetic anisotropy
energy is determined by the matrix elements of the spin-orbit interaction
between occupied states. Hence, the contributions from all the filled
bands must be considered to calculate the absolute value of magnetic
anisotropy. On the other hand, the magnetic damping is related to
the density of states at the Fermi level.

\begin{table}
\begin{centering}
\begin{tabular}{|c|c|c|c|c|c|c|}
\hline 
 & $4\pi M_{\textrm{eff}}$ (T) & $H_{2\parallel}$(Oe) & $H_{4\parallel}$(Oe) & $\alpha_{\perp}$ & $\alpha$($\varphi=0$) & $\alpha(\varphi=\pi/2)$\tabularnewline
\hline 
\hline 
S-LSMO & 0.3280$\pm$0.0011 & 37$\pm4$ & 0 & $(8.6\pm0.5)\times10^{-4}$ & $(8.4\pm0.3)\times10^{-4}$ & $(11\pm0.6)\times10^{-4}$\tabularnewline
\hline 
W-LSMO & 0.3620$\pm$0.0025 & 13.9$\pm0.9$ & 11.8$\pm1.2$ & $(4.7\pm0.7)\times10^{-4}$ & $(6.5\pm0.3)\times10^{-4}$ & $(5.3\pm0.3)\times10^{-4}$\tabularnewline
\hline 
\end{tabular}
\par\end{centering}

\centering{}\protect\caption{Summary of the parameters for S-LSMO and W-LSMO samples. \label{tab:SummaryValues}}
\end{table}

The damping term in the Landau-Lifshitz-Gilbert equation of motion
is $\frac{\alpha}{{\rm |}M{\rm |}}\left(M\times\frac{dM}{dt}\right)$,
therefore, anisotropy in damping can have two origins, one related
to the equilibrium orientation of magnetization $M$ (orientation
anisotropy) and the other depends on the instantaneous change in magnetization
$dM/dt$ (rotational anisotropy). In FMR experiments the magnetization
vector rotates around its equilibrium position, therefore, the rotational
anisotropy may be smeared out \citep{Seib2009}. The orientation anisotropy
is described by both interband and intraband scattering process. According
to Gilmore et al.\citep{Gilmore2010}, the latter is isotropic at
sufficiently high scattering rates at room temperature. We suspect
that the anisotropic damping in LSMO is due to its half-metallic band
structure. As a result of high spin polarization, interband scattering
is suppressed and the room temperature damping is dominated by intraband
scattering. The intraband contribution to damping exhibit anisotropy
for all scattering rates \citep{Gilmore2010} which agree well with
our experiments. The suppression of interband scattering is evidenced
by the ultra-low damping in the order of $10^{-4}$. Notably, the
absolute value of the observed anisotropy, 2.6$\times$10$^{-4}$
for S-LSMO and 1.2$\times$10$^{-4}$ for W-LSMO, is so small that
could not be identified reliably for a material with typical damping
values between 5$\times$10$^{-3}$ to 2$\times$10$^{-2}$.

In a microscopic picture, the Gilbert damping is proportional to the
square of SOC constant ($\xi$) and density of states at the Fermi
level, $\mbox{\ensuremath{\alpha\sim\xi^{2}D(E_{F})}}$. The shape
of the Fermi surface depends on the orientation of the magnetization
due to SOC. Hence, the anisotropy can be attributed to the angle dependence
of $D(E_{F})$ which is in turn induced by the substrate. The trend
reversal in the damping anisotropy of the two LSMO samples can be
explained by the modification of the Fermi surface and thus $D(E_{F})$
by strain relaxation. During the preparation of this paper, we noticed
a similar work in ultra-thin Fe layers deposited on GaAs substrate\citep{NatPhy2018}.
There, the anisotropy is attributed to interfacial SOC. This work
suggests that anisotropic damping can exist in bulk samples. 
\begin{acknowledgments}
The research is supported by the Singapore National Research Foundation
under CRP Award No. NRF-CRP10-2012-02. P. Yang is supported from SSLS
via NUS Core Support C-380-003-003-001. S.J.P is grateful to the National
University of Singapore for funding.
\end{acknowledgments}

\bibliographystyle{apsrev4-1}

\clearpage{}


\end{document}